 \documentclass[twocolumn,letterpaper]{igs}

  \usepackage{igsnatbib}
  \usepackage{stfloats}

    \usepackage[dvips]{graphicx}

\begin{document}

\title[Ross Ice Shelf RF attenuation]{Ross Ice Shelf \emph{in situ} radio-frequency ice attenuation}

\author[Barrella et al.]{Taylor BARRELLA,$^1$
  Steven BARWICK,$^2$ David SALTZBERG$^1$}

\affiliation{%
$^1$Department of Physics and Astronomy, University of California, Los Angeles, Los Angeles, California 90095, USA\\
$^2$Department of Physics and Astronomy, University of California, Irvine, Irvine, California 92697, USA}

\abstract{We have measured the \emph{in situ} average electric field attenuation length $\langle L_\alpha \rangle$ for radio-frequency signals broadcast vertically through the Ross Ice Shelf. We chose a location, Moore Embayment, south of Minna~Bluff, known for its high reflectivity at the ice-sea interface. We confirmed specular reflection and used the return pulses to measure the average attenuation length from 75--1250~MHz over the round-trip distance of 1155~m. We find $\langle L_\alpha \rangle$ to vary from $\sim$500~m at 75~MHz to $\sim$300~m at 1250~MHz, with an experimental uncertainty of 55 to 15~m. We discuss the implications for neutrino telescopes that use the radio technique and include the Ross Ice Shelf as part of their sensitive volume.}

\maketitle

\section{Introduction}

Several areas of investigation in Antarctica benefit from 
knowledge of the
attenuation length of radio-frequency transmissions in glacial ice.
These research areas
include radar mapping of subglacial
terrain and lakes,
location and characterization of subglacial streams, 
look-ahead radar for over-ice traverses,
and detection of
particle interactions in the ice.

The authors of this report
belong to the community of particle astrophysicists aiming to
identify high-energy neutrinos (produced in astrophysical sources) via
the radio emission accompanying their interaction in Antarctic
ice shelves and sheets~\cite{Kravchenko03, Gorham10, Allison09, Barwick07, Hoffman07}.
Neutrino
interactions initiate a cascade of charged particles moving in a
thin pancake approximately 0.1~m in extent that traverses tens of meters of
ice before extinction.   The charged particles move near the speed of
light, $c$, faster than the velocity of light in ice, which is reduced
by an index of refraction, $n$, of approximately 1.78.  The electric field
thus forms an electromagnetic shock wave analogous to a sonic boom, known
as Cherenkov radiation.  Given the size of the charged particle
pancake, the emission among all charges sums coherently for
frequencies up to several~GHz.  This coherence, and the possibility to
find high-energy neutrino-induced cascades in large natural media such
as ice, was first identified by \cite{Askaryan62}.

More recent numerical simulations confirmed and expanded the modeling
of this emission \cite{Halzen91, AlvarezMuniz03, Razzaque04}.
The process was
subsequently confirmed experimentally 
in a variety of dielectric media, including ice, using showers
initiated by high-energy electron and photon beams \cite{Gorham07}.

Because the neutrino detection rate is a strong function of the radio-transparency of
glacial ice, knowing the attenuation length to a reliable precision at
several locales in Antarctica is of key importance.
Extractions of attenuation length for a variety of Antarctic locations
has been ongoing for over several decades. For a recent compilation, see Fig.~1 in a previous article \cite{Barwick05}.
Most of these values are based on the
AC conductivity of ice (either directly measured or estimated from
impurity measurements), from which the loss tangent 
\begin{equation}
\tan \delta  = \epsilon '' /\epsilon '
\label{losstangent}
\end{equation}
and attenuation length are extracted using temperature profiles \cite{Fujita00, Matsuoka03, MacGregor07}.
In Eqn.~\ref{losstangent}, $\epsilon
'$ and $\epsilon ''$ are the real and imaginary parts of the
complex dielectric constant of the material, $\epsilon = \epsilon' - i\epsilon''$.  From the loss
tangent, and its expected variation with temperature,
the intensity loss in dB~m$^{-1}$, at a frequency $\nu$, is $8.686(2\pi\nu /2c)(\sqrt{\epsilon'}\tan\delta)$.  
This loss, known as the attenuation coefficient, $\alpha$, 
is related to the electric field attenuation length in meters by
\begin{equation}
L_\alpha = 1/\ln{\sqrt{10^{\alpha/10}}}.
\label{l_alpha}
\end{equation}
For a full description of the electromagnetic phenomena of radio
frequency attenuation in glacial ice see \cite{Bogorodsky85} and \cite{Dowdeswell04}.

Neutrino searches and the other fields discussed
should benefit 
from {\it in situ} measurements of the total attenuation length, to complement those of the AC conductivity.
Several
such recent measurements have been performed on the ice sheet of the Antarctic
plateau: one (250--1000~MHz) at Taylor Dome in East Antarctica near the Transantarctic
Mountains \cite{Besson08},
one (220--700~MHz) near the South Pole \cite{Barwick05}, one (140--160~MHz) at WAIS Divide \cite{Laird09},
and one using much lower frequencies ($\sim$5~MHz) at Siple Dome in West Antarctica \cite{Winebrenner03}.

In this article we report
the measurement of the attenuation length in a floating ice sheet over the range 75--1250~MHz.
Specifically, we performed a radar bounce through the Ross Ice Shelf at a location
in Moore Embayment just south of Minna Bluff 
(78$^\circ$45.022'~S, 164$^\circ$59.291'~E).  
This region was chosen because in previous work on basal ice shelf reflectivity 
it was found to be highly reflective to radar \cite{Neal79},
consistent with a smooth mirror-like surface below.  
Having a low-loss reflective
surface at the bottom allows us to take a measurement of the returned
power and directly extract an attenuation length.  The smoothness of the
lower surface of an ice shelf (Fimbul Ice Shelf) is confirmed by recent sonar measurements from the sea
below \cite{Nicholls06} to be mirror-like down to wavelengths of 0.0075~m. The wavelengths considered here, within the $n=1.78$ bulk ice, range from 0.13~m to 2.25~m, so we expect a high reflectivity. For this measurement we assume a nominal reflection coefficient of $R=1.0$. A true lower value would only lead to longer attenuation lengths from our data, and for our purpose, $R=1.0$ is conservative for neutrino detector design. We show later the effect of reasonable variations on $R$ and we find that the effect is small due to the presence of the square root and logarithm in Eqn.~\ref{l_alpha}.

Balloon-borne neutrino observations can include downward-pointing neutrino events whose radio emission is reflected back upward.
It should be noted that for a balloon observation of the entire ice shelf, not all locations will have excellent reflectivity \cite{Neal79}.
However, a fixed detector may be placed on the surface above the smoothest ice-sea interface. The thickness of Moore Embayment and its proximity
to McMurdo station hold promise for this location as a site for a
future neutrino detector deployed on its surface.
Although this ice is warmer and expected to have a shorter attenuation length than measured on the plateau, our simulations show that it will be sufficient to work as a neutrino detector, with an added boost from reflected events \cite{Barwick07}. The purpose of this work is to confirm that the attenuation length is at least as long as the value used in that study. Our results indicate attenuation lengths 1.5--2.5 times longer (better) than used in that estimate.

\section{Experimental Approach}

Our basic approach is similar to that of previous work \cite{Barwick05, Besson08}.   We transmit a high voltage impulse, with broadband frequency content, into the ice and measure the return power vs.~frequency in another antenna.   We compare that to the transmission of the same pulse through a short distance, $r$,  in air.  After accounting for the $1/r^2$ geometrical factor, the remaining loss is ascribed to attenuation, in which we include scattering. We assume 100\% reflection at the interface of ice and sea water, which yields a conservative value for the attenuation length.  Systematic uncertainties in the transmission and reception are reduced by using only voltage ratios.  Moreover, because we are measuring over a long distance, {\it i.e.} several times the attenuation length, the remaining systematic uncertainties in the ratio enter only through a logarithm, as shown below.

The transmitted power per unit area is reduced by two effects: a $1/r^2$ factor due to the spreading of the wave, and an exponential factor $\exp{(-2r/\langle L_\alpha \rangle)}$ due to losses, where $\langle L_\alpha \rangle$ is averaged over depth. The factor of 2 accounts for power being the square of the electric field.  We define $P_\nu$ to be the power spectral density at frequency $\nu$ of the pulse as received by a $50~\Omega$ receiver and $V_\nu$ to be $\sqrt{P_\nu \times 50\Omega}$.   We further define $V_{\nu,\mathrm{ice}}$ to be $V_\nu$ measured after being transmitted through the ice and back to our receiver, and $V_{\nu,\mathrm{air}}$ to be $V_\nu$ after being transmitted horizontally through the air between the two antennas separated by $d_\mathrm{air}$, typically 9~m. (For the in-air calibration, the antenna centers were 2~m above the top of the firn. Given the directivity of the antennas, the effect of multipath on the calibration is small.)  We show later that $d_\mathrm{ice}$ (twice the ice depth) is 1155~m.  Hence, before considering different transmission efficiency into air versus ice,
\begin{equation}
 V_{\nu,\mathrm{ice}}/V_{\nu,\mathrm{air}} = (d_\mathrm{air}/d_\mathrm{ice})\mathrm e^{-d_\mathrm{ice}/\langle L_\alpha\rangle}.
\label{ideal}
\end{equation}
As described below, we measured $T_\mathrm{ratio}$, the ratio of power the antenna transmitted into ice over that transmitted into air, accounting for an expected small difference due to impedance matching. For electric field measurements, we use $\sqrt{T_\mathrm{ratio}}$. Applying this correction twice, once for transmission and once for reception, Eqn.~\ref{ideal} becomes
\begin{equation}
 V_{\nu,\mathrm{ice}}/V_{\nu,\mathrm{air}} = \left(\sqrt{T_\mathrm{ratio}}\right)^2 \times (d_\mathrm{air}/d_\mathrm{ice})\mathrm e^{-d_\mathrm{ice}/\langle L_\alpha\rangle},
\label{withtratio}
\end{equation}
where $T_\mathrm{ratio}$ is a value between 0.8 and 1.2.

Solving for $\langle L_\alpha\rangle$ gives the equation we use to extract the attenuation length:
\begin{equation}
 \langle L_\alpha\rangle = d_\mathrm{ice}/\ln{\left(T_\mathrm{ratio} \frac{V_{\nu,\mathrm{air}}\,d_\mathrm{air}}{V_{\nu,\mathrm{ice}}\,d_\mathrm{ice}}\right)}.
\label{log1}
\end{equation}
Note that changes in the transmission into ice versus air, as well as any difference in beam pattern, enter only through a logarithm.

Care should be taken not to confuse average attenuation length $\langle L_\alpha \rangle$, which is averaged over depth, with $L_\alpha$. Some authors report $L_\alpha$ for a particular temperature and depth.

\begin{figure}
\centering{\includegraphics[width=86mm]{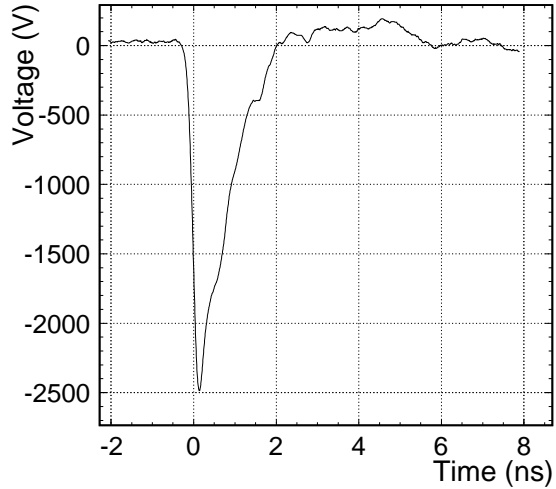}}
\caption{Unfiltered output of the high-voltage (HYPS) pulser into 50 $\Omega$.}
\label{pulser}
\end{figure}

\begin{figure}
\centering{\includegraphics[width=86mm]{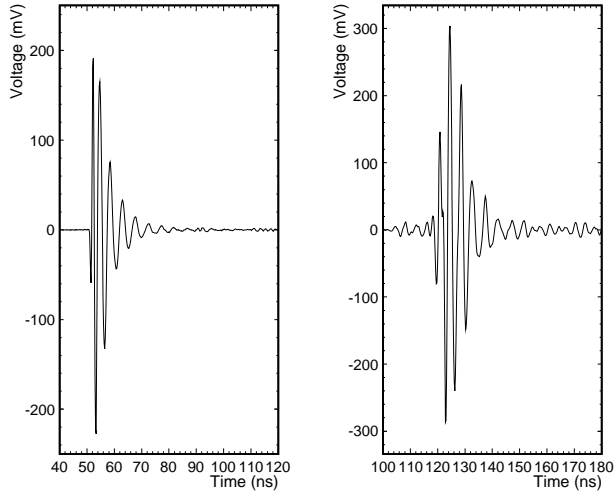}}
\caption{Typical waveforms as transmitted and recorded by the quad-ridged horns through 9~m of air (left) and 1155~m of ice (right). Both waveforms were recorded without the 900~MHz low-pass filter. Each waveform was attenuated or amplified to be approximately the same scale on the oscilloscope.}
\label{waveform}
\end{figure}

For our impulse we used a high-voltage pulser (HYPS, Grant Applied Physics) designed to drive a $50~\Omega$ load.  The output of the pulser is shown in Fig.~\ref{pulser} as measured with a 3~GHz bandwidth oscilloscope. It output a $-2.5$~kV peak with a 150~ps falltime.  We sent and received the impulses to the antennas via 23~m Heliax LDF4 $50~\Omega$ cables. On the receiving end an additional 4~m of LMR240UF cable was added. For most data, the signal was amplified by a Miteq low-noise amplifier model AFS3-00200120-10-1P-4-L with a gain of 49--54~dB from the highest to lowest frequencies. The filters discussed below were added after the amplifier and the signals were recorded by a Tektronix TDS684 oscilloscope.

\begin{figure}
\centering{\includegraphics[width=86mm]{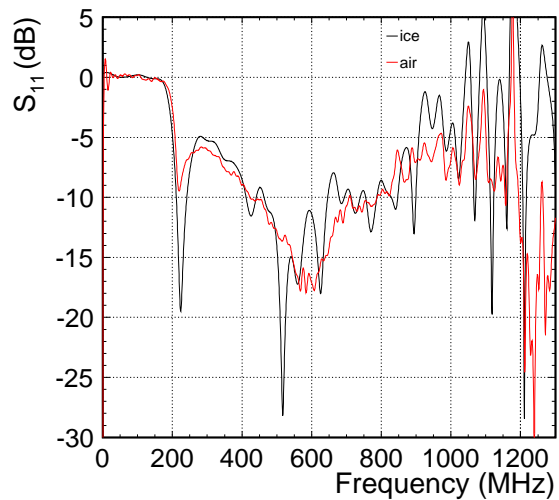}}
\caption{Fraction of incident power reflected back from the quad-ridged horn (relative to an open termination). The low-pass filter began to cut off signal past 900~MHz.}
\label{s11}
\end{figure}

\begin{figure}
\centering{\includegraphics[width=86mm]{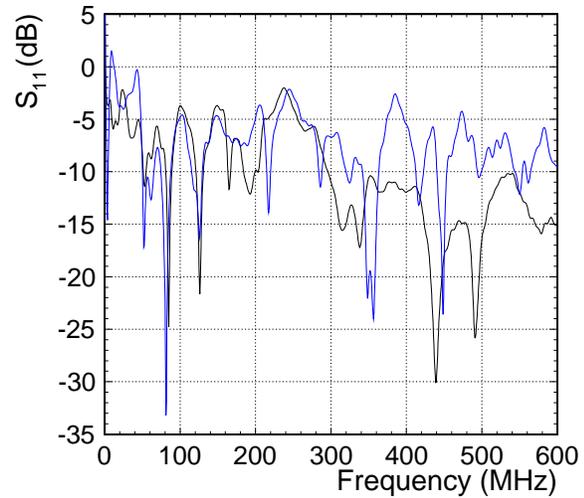}}
\caption{Fraction of incident power reflected back from the Yagi antennas (relative to an open termination) while in the snow. The two curves are for the two different antennas.}
\label{s11tv}
\end{figure}

Our main data were taken with quad-ridged horns built and designed by Seavey
Engineering Associates with a bandpass of 200 to 1280~MHz, specifically designed
for the Antarctic Impulsive Transient Antenna (ANITA) experiment \cite{Gorham10}.  Full details of the antenna are given in
\cite{Gorham09}. Typical waveforms transmitted and received through these antennas are shown in Fig.~\ref{waveform}. The preferential loss at high frequencies in ice is already apparent.

\begin{figure}
\centering{\includegraphics[width=86mm]{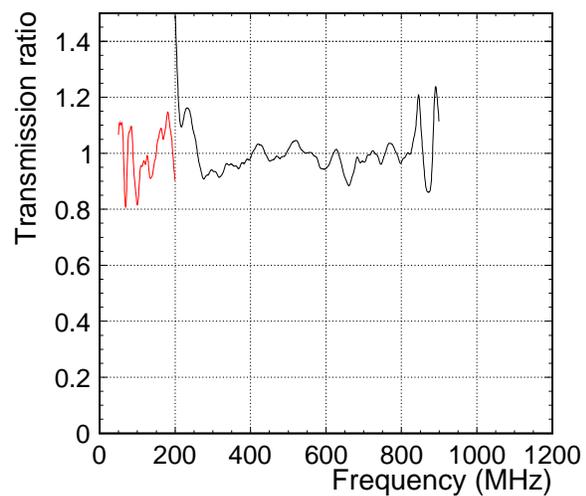}}
\caption{$T_\mathrm{ratio}$ for the quad-ridged horns (black) and Yagi antennas (red), with a 900~MHz low-pass filter. Beyond 900~MHz, we use $1.0\pm0.1$.}
\label{tratio}
\end{figure}

The transmission band of the antenna can be determined from the fraction of power sent to the antenna that is reflected back to the pulser, $S_{11}$.
The $S_{11}$ for the antenna pointing into air (as designed) is shown in Fig.~\ref{s11}.  Since the antenna was not designed for transmission directly into ice, we made $S_{11}$ measurements in various configurations facing down: directly on the snow surface, 1~m above the snow surface, and buried 0.5~m below the snow surface. We also compared these to having the antenna face up to the sky.
Little difference was found among these, and we show $S_{11}$ in Fig.~\ref{s11} for the nominal transmission into ice.  For lower frequencies we used Yagi antennas whose $S_{11}$ are shown in Fig.~\ref{s11tv}.  The critical parameter, however, is the ratio of transmitted power:
\begin{equation}
T_\mathrm{ratio}\equiv \frac{1-10^{S_{11,\mathrm{ice}}/10}}{1-10^{S_{11,\mathrm{air}}/10}},
\end{equation}
where $S_{11}$ is measured in dB.  The values of $T_\mathrm{ratio}$ versus frequency are shown for the quad-ridged horn and Yagi antennas in Fig.~\ref{tratio}. Transmission inefficiencies such as Ohmic losses in the antenna cancel in the ratios used.

To calculate the distance the radio emissions travel through ice, $d_\mathrm{ice}$, we measure the two-way transit time between the transmitted and received pulse and use a model for index of refraction, $n$, versus depth, $z$.  We measured the total propagation time,
$\Delta t=6783$~ns, to a precision of 10~ns.
We follow
\cite{Dowdeswell04} in modeling the ice as two regions: a slab
of bulk ice with constant $n$ surmounted by a firn layer with varying
$n(z)$.  For consistency with our previous work we take the bulk as
$n=1.78\pm0.03$, where the uncertainty comes from the range of values
summarized in~\cite{Bogorodsky85}.  

We use the Schytt model \cite{Schytt58}
of the firn layer's
index of refraction, $n$, versus density, $\rho$:
\begin{equation}
n(z)=1.0+0.86 \, \rho(z),
\end{equation}
where $\rho(z)$ is the specific gravity we measured using core samples on the Ross Ice Shelf (Williams Field):
\begin{equation}
\rho(z) = 1.0 - 0.638\mathrm \, e^{-z/34.7 \, \mathrm{m}}.
\end{equation}
In the model, the index of refraction of the firn matches the deep ice at $z=67$~m, consistent with the results in Fig.~2 of \cite{Dowdeswell04}, beyond which we take $n(z)=1.78$.    The value of $d_{ice}$  can then be found by integrating over the depth:
\begin{equation}
\Delta t = 6783~\mathrm{ns} = \frac 2 c \int_0^{d_\mathrm{ice}/2}{n(z)\,\mathrm d z}.
\end{equation}
Even extreme variations in the firn layer model, {\it e.g.} linear vs.~depth, are negligible ($<$5~m). The dominant uncertainty comes from the choice of $n$ for the bulk ice ($\pm0.03$), based on the distribution of values from \cite{Bogorodsky85}.  For our site we measure $d_\mathrm{ice}$ = $1155\pm20$~m.  This corresponds to an ice depth of $577.5\pm10$~m, agreeing with a measurement of $572\pm6$~m made by another team with a different method $\sim$1~km away~\cite{Gerhardt10}. Their smaller uncertainty corresponds to a smaller uncertainty used on the index of refraction.

We derive $\langle L_\alpha \rangle$ from the data using Eqn.~\ref{log1}.
The value of $V_\nu$ includes the power received in the co- and cross-polarizations relative to the transmitted signal.  When plotting the data, we binned the data points with intervals of 25~MHz, averaging $P_{\nu,\mathrm{ice}}$ and $P_{\nu,\mathrm{air}}$ over the width of the bin.  
The variation within 25~MHz was small, but the standard deviation divided by the square root of the number of bin elements is included in the error bars.  To calculate $\langle L_\alpha \rangle$, we averaged $T_\mathrm{ratio}$ over the width of the bin, although its variation across the bin is negligible.

\begin{figure}
\centering{\includegraphics[width=86mm]{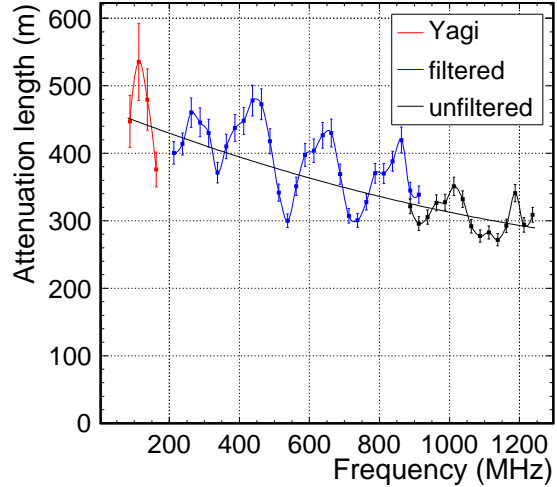}}
\caption{Calculated average attenuation length, $\langle L_\alpha\rangle$, as a function of frequency. The red line corresponds to data taken with the Yagi antennas, and the blue and black lines correspond to quad-ridged horn data taken with and without a low-pass 900~MHz filter. The modulation vs.~frequency is probably an artifact, as described in the text.}
\label{main}
\end{figure}

The final results are shown in Fig.~\ref{main}.  The red data are from the Yagi over its frequency band, which was indicated by a repeatable and small $S_{11}$.  The blue data are from the quad-ridged horns with a low-pass filter at 900~MHz (Minicircuits NLP-1000).  The black curve shows a continuation of data taken to higher frequencies without the filter up to 1250~MHz.  We discuss systematic uncertainties and several issues with the data below.

In principle the Yagi data should have also given results up to 600~MHz, where we still saw returned power.  The $T_\mathrm{ratio}$ value looked stable to within 20\% for this data.   However, the $S_{11}$ in ice indicated that up to half the power could be reflected and we determined the Yagi data to be unreliable above 200~MHz.  Analyzing that data yielded attenuation lengths between 200 and 600~MHz about 75~m shorter than the nominal quad-ridged horn data.

\begin{figure}
\centering{\includegraphics[width=86mm]{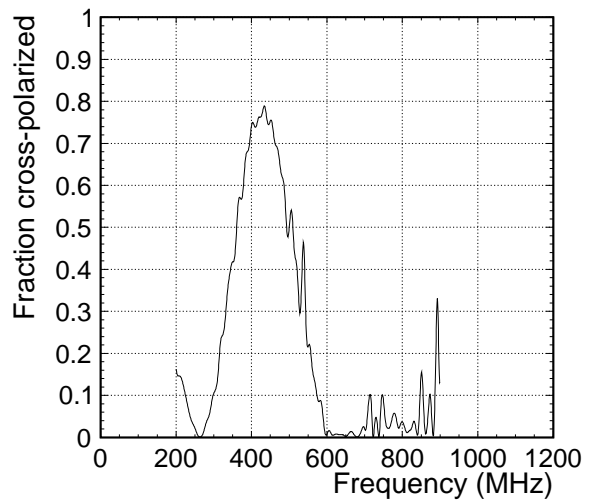}}
\caption{Fraction of the received signal power that was cross-polarized. These data were taken by rotating the receiver by 90$^\circ$ after taking the filtered, co-polarized quad-ridged horn data.}
\label{cross}
\end{figure}

At some frequencies we measured a significant amount of power arriving in the cross-polarization relative to the transmitted signal.  As shown in Fig.~\ref{cross}, this fraction had a peak at about 80\% at 450~MHz, which was only seen in ice data, not air.
This effect could be due either to a rotation of polarization in transit through the ice or due to local resonant interaction of the antenna with the ice.   In either case, keeping the received power in the calculation is correct, given the definition of $\langle L_\alpha \rangle$.
We are sure the signal traveled through the full depth of ice because of its arrival time.  No significant delay between the two polarizations was apparent.  The position and magnitude of the peak in Fig.~\ref{cross} did not depend on the orientation of the antennas relative to the ice.

Because of the functional form of Eqn.~\ref{log1}, the main known source of systematic error comes from uncertainty in $d_\mathrm{ice}$, 20~m out of 1155~m, arising mostly from the uncertainty of $n$ in the ice. Other sources of error include a 10\% uncertainty in $T_\mathrm{ratio}$ and the variation of power within each 25~MHz bin.

Our assumption of perfect reflectivity, {\it i.e.} reflection coefficient $R=1.0$, is conservative in that any lower value would yield longer attenuation lengths in the analysis. That is, we would detect more neutrino events.  At our site, the reflectivity is high~\cite{Neal79}, no worse than $R=0.25$ and likely much higher.  For completeness we show the effect of different assumed values for $R$ in Table~\ref{rtable}.  Because of the square root and logarithm in Eqn.~\ref{l_alpha}, the variation is modest.  \cite{Neal79} calculates a theoretical reflection loss at the seawater/glacial ice boundary of $-0.77$~dB ($-0.8$~dB if frozen sea ice is present) which corresponds to $R=0.83$.  Had we used that value, the effect on our answers would have been longer by typically a few percent and never more than 10\%.

\begin{table}
\begin{tabular}{c||c|c|c|c}
\hline
\hline
~~~~~~~~~~~~~~~~~~~~~~ & \multicolumn{4}{|c|}{$\nu$ (MHz)} \\
$R$ & 112.5 & 412.5 & 812.5 & 1212.5 \\
\hline
1.25 & 509. & 429. & 357. & 286. \\
1.00 & 535. & 448. & 370. & 294. \\
0.75 & 574. & 474. & 388. & 305. \\
0.50 & 638. & 518. & 416. & 322. \\
0.25 & 789. & 613. & 476. & 357. \\
\hline
\hline

\end{tabular}

\caption{Variation of the average extracted attenuation length (in m), $\langle L_\alpha \rangle$, as the assumed power reflection coefficient, $R$, is varied from its nominal value of 1.0. Note that $\langle L_\alpha \rangle$ only increases for $R$ smaller than the nominal value.}
\label{rtable}
\end{table}

An additional uncertainty, probably dominant, is apparent in Fig.~\ref{main} as a modulation versus frequency of about 183~MHz.  Tests in the field showed that this modulation depended on the local interaction of the antenna with the surrounding dielectric snow.  The modulation frequency depended on the depth at which the antenna was buried or how far above the ice it was tested.  It also depended on whether the horn was filled with snow or just air.  However, the modulation could not be reduced.  Another possibility is that the modulation was due to birefringence, by superposing two signals arriving at slightly different times due to different propagation speeds along the crystal axes. A modulation as observed is consistent with a 0.1\% difference in wave speeds due to birefringence, as has been observed at the South Pole \cite{Besson10, Matsuoka09}.  Such an effect could also be mimicked by partial reflection at or within intermediate layers or near the bottom. If due to a single layer in the bulk ice, the 183~MHz would correspond to a thickness of 0.46~m. Unfortunately, we did not take enough data at different angles to confirm or rule out these effects.
Rather than smoothing out the data, we present the results at each frequency bin and assign an uncertainty of approximately 55~m at low frequencies and 15~m at high frequencies.  A three-parameter fit is given below that effectively averages out these variations.

\section{Summary}

We have transmitted and received high-voltage impulses from
directional horn antennas through the Ross Ice Shelf to the sea water
below and back.  We have demonstrated that the reflection at the ice-sea
interface is consistent with an excellent mirror at these frequencies.
In addition, we have extracted the average attenuation length over a
range of VHF and UHF frequencies (75~to 1250~MHz) and found it to
range from 500~m to 300~m with an uncertainty of order the RMS of the
variation, or 55~m to 15~m over this range.  Since the variations are
small, if we assume they are an artifact, our E-field attenuation
length (in~m) can be summarized with a parameterization over the range 75~to
1250~MHz as
\begin{equation}
\langle L_\alpha \rangle = 469 - 0.205\,\nu + 4.87\times 10^{-5}\,\nu^2,
\label{resultsfit}
\end{equation}
where $\nu$ is in MHz. The use of a quadratic fit is for convenience and not motivated by a particular physical model. The fit values, with uncertainties, on the terms in Eqn.~\ref{resultsfit} are $469\pm13$, $-0.205\pm0.036$, and $(4.87 \pm 2.34) \times 10^{-5}$. 
Although the data are not expected to be described by a constant loss tangent, $\tan \delta$, the attenuation length decreases with frequency.  Naively, the data in this range can be summarized by $\tan \delta \sim 0.0003$, corresponding to a Q-factor $Q \equiv 1/\tan \delta \sim 3300$.
Few direct measurements of the attenuation length have been made over
these frequencies and our results can be used to constrain modeling based on
impurities and temperature.

Our measurements were systematically limited by our understanding of
birefringence and interaction of the antennas with the ice.  Based on
our experience, we recommend future measurements should take data at
numerous angles with respect to the ice fabric. For example, see data taken at Upstream B Camp~\cite{Liu94}, results which warn that the crystal fabric can change orientation on scales of 100~m.  Care should also be
taken in antenna design or simulation to understand the interaction
with the dielectric medium.   The impulses were strong enough that
reflected signals from intermediate layers were visible, so a mapping
of birefringence vs.~depth should be possible if many angles are taken.

These attenuation lengths are long enough, comparable to the depth,
that neutrino detectors proposed for this region of the Ross Ice Shelf
appear viable.  In addition, balloon flights that have observed this region are also sensitive to neutrino interactions. The coherence of the pulse is well maintained with no
distortion beyond that due to the attenuation.  The lengths are
conservative in that if the mirror at the ice-sea interface is not
perfect, the actual attenuation lengths would be longer.

\section{Acknowledgments}

We are indebted for the logistical support provided by Raytheon Polar Services and the National Science Foundation, and particularly to Kevin Emery for field support. We thank Jordan Hanson for finding an error in Fig.~\ref{cross}, which we have fixed. This material is based upon work supported by the US National Science Foundation's Office of Polar Programs and Department of Energy's Office of Science.



\end{document}